\documentstyle[multicol,epsf,aps]{revtex}
\begin{document}
\draft

\title{Band termination in the {\boldmath $N=Z$} Odd-Odd Nucleus $^{46}$V}

\author{
S.~M.~Lenzi,$^{1}$
D.~R.~Napoli,$^{2}$
C.~A.~Ur,$^{1,3}$
D.~Bazzacco,$^{1}$
F.~Brandolini,$^{1}$
J.~A.~Cameron,$^{4}$
E. Caurier,$^{5}$
G.~de~Angelis,$^{2}$
M.~De~Poli,$^{2}$
E.~Farnea,$^{2}$
A.~Gadea,$^{2}$
S.~Hankonen,$^{6}$
S.~Lunardi,$^{2}$
G.~Mart\'{\i}nez-Pinedo,$^{7}$
Zs. Podolyak,$^{2}$
A. Poves,$^{8}$
C.~Rossi~Alvarez,$^{1}$
J. S\'anchez-Solano,$^{8}$
and H.~Somacal$^{2}$}

\address{
$^{1}$Dipartimento di Fisica and INFN, Padova, Italy\\
$^{2}$Laboratori Nazionali di Legnaro, INFN, Legnaro, Italy\\
$^{3}$H. Hulubei National Institute of Physics and Nuclear Engineering, 
Bucharest, Romania\\
$^{4}$Department of Physics and Astronomy, McMaster University, Hamilton, 
Canada\\
$^{5}$Physique Th\'eorique, CRN, Strasbourg, France\\
$^{6}$Department of Physics, Jyv\"askyl\"a\\
$^{7}$University of Aarhus, Aarhus, Denmark\\
$^{8}$Departamento de F\'{\i}sica Te\'{o}rica C-XI, Universidad Aut\'{o}noma de 
Madrid, E-28049 Madrid, Spain\\
}

\date{May 5, 1999}
\maketitle

\begin{abstract}
  High spin states in the odd-odd $N=Z$ nucleus $^{46}$V have been
  identified.  At low spin, the $T=1$ isobaric analogue states of
  $^{46}$Ti are established up to $I^{\pi} = 6^+$. Other high spin
  states, including the band terminating state, are tentatively
  assigned to the same $T=1$ band.  The $T=0$ band built on the
  low-lying $3^+$ isomer is observed up to the $1f_{7\over2}$-shell
  termination at $I^{\pi} = 15^+$.  Both signatures of a negative
  parity $T=0$ band are observed up to the terminating states at
  $I^{\pi} = 16^-$ and $I^{\pi} = 17^-$, respectively. The structure
  of this band is interpreted as a particle-hole excitation from the
  $1d_{3\over2}$ shell. Spherical shell model calculations are found
  to be in excellent agreement with the experimental results.
\end{abstract}

\begin{multicols}{2}
  
In the last few years, the study of $1f_{7\over2}$-shell nuclei has
gained renewed interest due to dramatic advances in experimental and
theoretical techniques.  Close to the middle of the $f_{7\over2}$
shell, nuclei show strong deformation near the ground state.  At high
spin, the interplay between single particle and collective behaviour
is manifested in backbending and band termination phenomena. The
$N=Z=24$ nucleus $^{48}$Cr has the maximum number of valence particles
to develop deformation and is considered the best rotor in this shell
with a deformation parameter of $\beta \approx
0.26$~\cite{48Cr,Cau95}.  As one moves away from $^{48}$Cr, the
collective behaviour is less pronounced, shell effects are more
evident and nuclei evolve towards a spherical shape. The
$f_{7\over2}$-shell nuclei thus provide an excellent opportunity to
study the interplay between single particle and collective degrees of
freedom as a function of both angular momentum and valence particle
number.

Another interesting feature that can be studied in these nuclei as a
function of spin and mass number is the pairing interaction. While
like-nucleon pairing is of $T=1$ character, proton-neutron pairs can
be coupled to $T=0$ and $T=1$. It has been argued that the
contribution of both modes of $pn$ pairing decreases rapidly with
increasing $|N-Z|$~\cite{Lan97}.  Of particular interest are
therefore, in this respect, those nuclei which lie on the $N=Z$ line.
In the last few years, apart from $^{48}$Cr other even-even $N=Z$
nuclei have been studied up to high spin in this shell, i.e.,
$^{52}$Fe~\cite{52Fe} and very recently $^{44}$Ti~\cite{Snec98}.
Band-like structures built on the $T=0$ ground states have been
reported in all cases. The competition between $T=0$ and $T=1$ pairing
modes, however, can be better studied in the odd-odd $N=Z$ nuclei of
the $f_{7\over2}$ shell, in which ground states have quantum numbers
$T=1$ $I^{\pi}=0^+$, but the $T=0$ structures appear at low excitation
energies and rapidly become yrast.

The pairing interaction is well known to decrease with increasing spin
as the mechanism of generating angular momentum by breaking pairs of
valence particles and aligning their spins along the rotational axis
could become energetically favored.  When all valence particles are
aligned with the rotational axis, band terminating states are built.
Experimentally, non-collective band-terminating states of natural
parity have been reported recently in several nuclei in this
shell~\cite{48Cr,Snec98}.  Moreover, non-natural parity bands have
been observed in many $f_{7\over2}$ shell nuclei which correspond to
larger deformed structures than those of natural parity~\cite{Pol}.
Despite the many recent advances, however, there remains a deficiency
in the observation of unnatural parity bands. In many cases the
terminating states have not been identified, a fact which becomes
crucial for the interpretation of the intrinsic structure.

The most deformed odd-odd self-conjugate nuclei in the $f_{7\over2}$
shell are $^{46}$V and $^{50}$Mn, with one proton-neutron pair
subtracted or added to $^{48}$Cr, respectively. In a recent work, high
spin states in $^{50}$Mn have been reported for the first
time~\cite{50Mn}. The g.s. $T=1$ band, isobaric analogue of the g.s.
band in $^{50}$Cr, has been reported up to the $4^+$ state, while the
$T=0$ yrast band has been observed up to the band terminating $15^+$
state.

In this work we report the first observation of high spin states in
the odd-odd $N=Z$ nucleus $^{46}$V.  The $T=1$ isobaric analogue
states of $^{46}$Ti are established up to spin $6^+$. The
energetically favoured $T=0$ band built on the isomeric $3^+$ state is
extended up to its $f_{7\over2}$-shell terminating 15$^+$ state.  Both
signatures of another $T=0$ band of negative parity are observed up to
the band termination.  The intrinsic structure of this band is
interpreted as $(d_{3\over2})^{-1} (pf)^7$ configuration.  This is the
first time that both signatures of an unnatural parity band are
observed up to the band termination in this mass region.  Very few
states were known previously in $^{46}$V.  In particular, the highest
state tentatively reported by Poletti {\it et al.}~\cite{Pol81} was
the $J=9^+$ at 3094 keV which we confirm. At the same time of this
work, other two groups have studied $^{46}$V, at low~\cite{46VvB} and
high spins~\cite{46VOL}. We agree in general with their observations
and report several more new levels and transitions to the scheme of
$^{46}$V which are crucial for the interpretation of the underlying
structures.

High spin states in $^{46}$V have been populated via the
$^{28}$Si($^{24}$Mg,$\alpha pn$) reaction, with a 100 MeV $^{24}$Mg
beam provided by the XTU Tandem Accelerator of the Legnaro National
Laboratory.  The target consisted of $0.4 mg/cm^2$ of $^{28}$Si
(enriched to 99.9\%).  Gamma rays were detected with the GASP array,
comprising 40 Compton-suppressed HPGe detectors and an 80-element BGO
ball which acts as a $\gamma$-ray multiplicity and sum-energy filter.
Light charged particles were detected with the ISIS array, consisting
of 40 ($\Delta E, E$) Si telescopes. Gain matching and efficiency
calibration of the Ge detectors were performed using $^{152}$Eu,
$^{56}$Co and $^{60}$Co radioactive sources.

The level scheme of $^{46}$V deduced from the present work is shown in Fig. 1. 
It has been built on the basis of a 
$\gamma$-$\gamma$-$\gamma$-coincidence 
cube and a $\gamma$-$\gamma$-matrix constructed from all events in which 
one proton and one $\alpha$-particle were detected in the ISIS array. In this
latter matrix, the kinematical Doppler correction has been performed according
to the detection geometry of the charged particles. Gamma-ray spectra obtained
from setting coincidence gates in this matrix are shown in Fig. 2.
The spin-parity assignments were deduced from the $\gamma$-ray angular 
distribution, the directional 
correlation from oriented states (DCO) ratios and from the decay 
pattern. 

The most strongly populated structure in $^{46}$V is the positive 
parity $T=0$ band A built on the low-lying $3^+$ isomeric state 
($T_{1\over2}$=1.04 ms). 
This band is now extended 
up to the $15^+$ terminating state, the maximum spin available to $^{46}$V in a 
pure $f_{7\over2}$-shell configuration.  
As happens in $^{50}$Mn, where the $T=0$ bandhead has a spin $J=5$, 
while the $pn$ interaction favours in energy the 
$J=1$ and $J=7$ couplings, the strong quadrupole field near the
middle of the shell gives rise to a $K^{\pi}=3^+$ bandhead for the $T=0$ yrast 
band. In the
Nilsson scheme this corresponds to a proton and a neutron in the 
$[301]{3\over2}$ orbit.
The low spin behaviour of the band, including the state $J=1^+$ at 994 keV, 
can be interpreted
as the smooth alignment of the last proton and neutron in a $T=0$ coupling.

Bands B and C are also identified as $T=0$ structures as they 
have no counterparts in $^{46}$Ti. The DCO analysis indicates that states in
bands B and C decay by electric quadrupole
transitions to the states in band A, which is consistent with the fact that 
magnetic dipole transitions between 
$T=0$ states are hindered.

Above the band termination, a very weak transition feeding the 15$^+$ state is 
reported. 
Its intensity is two orders of magnitude weaker than the 
$15^+\rightarrow13^+ $ transition
which confirms the terminating character of the $15^+$ state. To build a 
spin greater than $J=15$ one proton or one neutron has to be excited to 
the $f_{5\over2}$ subshell which lies about 6 MeV above the $f_{7\over2}$. On 
the basis 
of DCO analysis we assign $J^{\pi}=16^+$ to the state
at 11.753 MeV.

The $0^+$, $2^+$, $4^+$ and $6^+$ states of band D in $^{46}$V at 
excitation energies 
of 0, 915, 2054 and 3365 keV are interpreted as the $T=1$ isobaric analogues of
the 0, 889, 2010 and 3300 keV states of $^{46}$Ti. 
These $T=1$ states were populated very weakly 
in the present experiment. They decay preferentially by M1 transitions to
$T=0$ states. Other high spin states which
decay to band A are tentatively assigned as $T=1$ states. For the decay of
the $(8^+)$ and $(11^+)$ states at 4843 and 7725 keV, the DCO
analysis was not conclusive. For the 3702 and 1617 keV transitions 
from the $12^{(+)}$ and $14^{(+)}$ states, respectively, 
the DCO ratios suggest a dipole character. 
The tentative parity assignments are based on the decay pattern and  
the comparison with the level scheme of $^{46}$Ti. As will be discussed
below, shell model calculations are in general consistent with this assignment.

On the left side of Fig. 1, two very regular band structures are drawn which 
are interpreted as the two signatures of a negative parity $T=0$ band. 
The DCO analysis of the decay transitions of 333, 372, 750, 1198 and 1698 keV
gives ratios consistent with dipole transitions, while for the 451 keV
line the obtained DCO ratio suggests a dipole transition of $\Delta 
J=0$ type.
The parity assignment, based on the decay pattern and on systematics, is 
also sustained by shell model calculations. Non-natural 
parity bands in other $f_{7\over2}$ nuclei can be described as a 
$f_{7\over2} d_{3\over2}^{-1}$ particle-hole excitation. In this picture, the
terminating states of both $\alpha=0$ and $\alpha=1$ signatures should be 
$J=16$ and $J=17$, respectively, in agreement with what we observe. As stated 
above, intruder bands of
this type have been already observed in odd and even-even nuclei in the
$f_{7\over2}$ 
shell, but it is the first time that such a band is 
observed up to the band termination in both signatures. 

We have observed a weak 128-keV transition from the $2^-$ 
state to a level at 1236 keV which further decays to the $1^+$ level. These two 
transitions are very weak and therefore no angular correlation analysis was  
possible. On the basis of the decay pattern and the shell model predictions we 
tentatively assigned $J^{\pi}=0^-$ to the level at 1236 keV.
                                                                      
Both E and F bands are much more regular than the 
positive parity $T=0$ band, suggesting more collectivity. A backbending 
at spin $J\simeq10$ is evident from Fig. 1. 
Up to the backbending there is not signature splitting, 
as can be seen on the inset to Fig. 3b.
There, the experimental excitation energy normalized to that of a rigid rotor 
($E-0.05473
J(J+1)$) for each signature is plotted. At spins higher than $J=8$, the slopes 
of the curves change, indicating a change of configuration. 
It can also be observed that both signatures terminate in a favoured 
way~\cite{AR}.  
The $T=0$ nature of the band hinders the
magnetic dipole transitions between two signature partner states. 

On the right side of Fig. 1, other levels are reported as belonging to a negative
parity band G. The tentatively assigned negative parity is based on DCO
analysis and decay pattern. 
  
As stated above, SM calculations in the full $pf$ shell space have been shown 
to give a very good description of $f_{7\over2}$-shell nuclei. We 
have performed these calculations to study the positive parity states in 
$^{46}$V with the 
code ANTOINE~\cite{ANTOINE} using the KB3
interaction and the experimental single-particle energies of $^{41}$Ca. 
The theoretical spectrum for the $T=0$ yrast band and the $T=1$ g.s. band of 
$^{46}$V obtained from these calculations is
compared with the experimental data in Fig. 3a. The agreement is very good.
The B(E2) and B(M1) values for some levels of both $T=1$ and $T=0$ positive 
parity bands are reported in Table I. Some
significative experimental and theoretical branching ratios (BR) are also 
quoted. The theoretical BR were calculated using the experimental 
transition energies. The decay properties of both bands are in good 
agreement with the experimental results.   

In the SM framework, the contribution to the energy of each spin state 
of the different pairing terms can be calculated~\cite{PMP}. We have 
calculated the expectation
value of the pairing correlation energy in perturbation theory, as described in
ref.~\cite{MP46V} for the $T=1$ isobaric analogue states 
in $^{46}$V and $^{46}$Ti. The results are reported in Fig. 4. In the
upper part of the figure, the contributions of the $T=1$ pairing terms are
plotted for $^{46}$V. The most important contribution arises from the $pn$
$T=1$ pairing which decreases with increasing angular momentum. At $J=8$ this
pairing mode becomes very small, which is consistent with the fact that the
first pair which aligns with the rotational axis is a $pn$ pair. On the other
hand, the contribution of the like-nucleon pairing remains almost constant.
The opposite behaviour is found for $^{46}$Ti, as shown in the middle of 
Fig. 4.  
The number of valence neutrons in $^{46}$Ti is twice that of the protons.
Therefore, at $J=0$ the $nn$ pairing contribution is twice that of the protons.
 At $J=2$ the $nn$ pairing strength decreases but also the 
$pp$ contribution does, although its change is not so marked. At $J=4$ 
both contributions become almost equal but at $J=6$, while the $nn$ term 
remains constant, the $pp$ contribution decreases by $\sim$30\%. 
This picture would suggest that there is a smooth alignment of both pairs of 
protons and neutrons with increasing spin. 
On the other hand, the $pn$ $T=1$ pairing strength in $^{46}$Ti follows the 
same behaviour of the like-nucleon pairs in $^{46}$V. 
In the bottom of Fig. 4, the contributions of
both $T=0$ and $T=1$ ($pp, nn$ and $pn$) pairing terms are given for $^{46}$V. 
At low spin, the $T=1$ pairing dominates over the $T=0$. 
However, by comparing with the $pn$
contribution in the upper panel, one can conclude that above $J=2$ the 
$T=0$ $pn$ pairing is more important than $T=1$ $pn$ pairing. 

The excitation energies at each spin of the $T=1$ analogue states
in $^{46}$V are greater than those in $^{46}$Ti up to $J=6$. 
This energy difference could be interpreted as a Coulomb effect due to the 
fact that when a proton pair aligns in $^{46}$Ti, the Coulomb energy 
decreases, as has been pointed out in ref.~\cite{OLe97}. On the contrary, 
the alignment of $pn$ pairs would not affect the Coulomb energy in $^{46}$V. 
This energy difference is also reproduced by SM calculations. However, 
SM calculations predict that the excitation energy of the $J=8$ state in 
$^{46}$V should be {\it smaller} than 
its analogue in $^{46}$Ti. This fact is consistent with the data but an
interpretation in terms of alignments is not straightforward from Fig.4.

For the highest spin states of band D we have also considered the 
possibility of a negative parity character. For this purpose and to study the
regular negative parity bands E and F we have also
performed SM calculations with the code ANTOINE in the full $sd$
and $pf$ valence space allowing for one hole in the $d_{3\over2}$ orbit as in
ref.~\cite{PSS}.
The SM calculations suggest that if the high spin
states in band D were of negative parity, they should decay to bands E, F or G.
We were not able, however, to observe such transitions. Further measurements
are needed to give a definitive assignment to these levels.
 
In a
self-conjugate nucleus such as $^{46}$V, a particle-hole excitation can be 
thought as a proton hole coupled to the g.s. band in 
$^{47}$Cr and a neutron hole
coupled to that in $^{47}$V. Since these two nuclei have $J^{\pi}={3\over2}^-$ 
ground states, 
one would expect $J^{\pi}=0^-$  or $J^{\pi}=3^-$ spins for the bandhead of 
the negative parity bands in $^{46}$V. Shell model 
calculations predict that the first three states ($0^-, 1^-, 2^-$) 
are almost degenerate. This can be understood by taking
into account that the $1^-$ and $2^-$ states can also be built by coupling 
the $d_{3\over2}$-hole 
with the states $J^{\pi}={5\over2}^-$ and $J^{\pi}={7\over2}^-$ 
in $^{47}$V or $^{47}$Cr,
which are also almost degenerate with the g.s. $J^{\pi}={3\over2}^-$.
Finally, the backbending observed at $J=10$ in $^{46}$V is consistent with 
the backbending in the $A=47$ mirrors at $J={17\over2}$. Data are compared with
SM calculations in Fig. 3b.

The SM predictions for the reduced transition amplitudes in bands E and F are
in agreement with the experimental decay patterns. The branching ratios for 
electric quadrupole transitions connecting states with $\Delta J=1$ are 
very small in comparison with those connecting $\Delta J=2$ states. Moreover, 
the B(M1) values are less than $5\times10^{-3}$ $\mu^2_N$ in all cases. 
The intrinsic quadrupole moments derived from the
calculated static quadrupole moments and the $B(E2)$ values are very similar at
low spin and decrease very slowly up to the backbending, 
indicating rotational behaviour~\cite{Cau95}. At high spin, the $B(E2)$ values 
are consistent with a terminating band process. 
Both signatures terminate at the maximum spin which can be built in the 
$(f_{7\over2})^7 (d_{3\over2})^{-1}$ configuration. 

In summary, the level scheme of $^{46}$V has been greatly extended. 
Positive and negative parity $T=0$ band structures have been identified 
and observed for the first time up to band termination. 
In addition, the isobaric analogue states of the $T=1$ band in
$^{46}$Ti are established to $J^{\pi}=6^+$. Higher spin states, including the
band terminating $J^{\pi}=14^+$, which decay to
the $T=0$ band A are tentatively assigned to the $T=1$ band.
Excellent agreement is found between the experimental data and calculations. 
Pairing properties of each isospin channel are studied  
as a function of angular momentum.
 
We are grateful to A. V. Afanasjev, M. A. Nagarajan, A. Vitturi and R. Wyss 
for fruitful discussions.

\end{multicols}

\begin{multicols}{2}

\begin{figure}
  \epsfxsize=0.9\columnwidth
  \epsffile{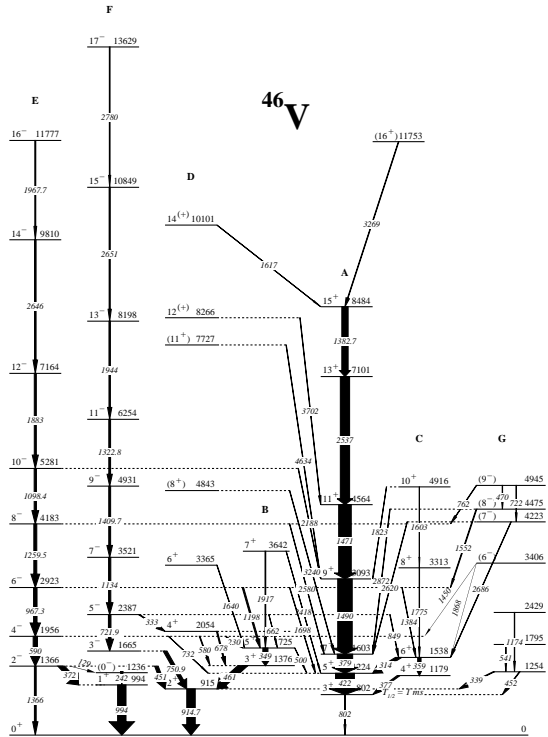}
  \caption{Level scheme of $^{46}$V from the present work.  Transition
    and level energies are given in keV.}
  \label{scheme}
\end{figure}

\begin{figure}
\epsfxsize=0.9\columnwidth
\epsffile{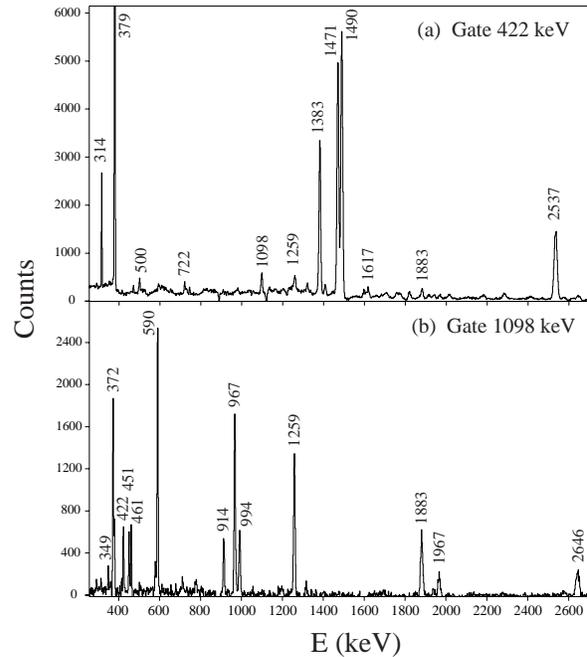}
\caption{$\gamma$-ray spectra generated by setting coincidence gates in the 
  $\gamma$-$\gamma$ kinematically Doppler corrected matrix.  (a) The
  spectrum in coincidence with the 422 keV transition in the positive
  parity $T=0$ band A. (b) The spectrum in coincidence with the
  1098-keV transition in the negative parity band E.}
\label{spectra}
\end{figure}

\begin{figure}
\epsfxsize=0.9\columnwidth
\epsffile{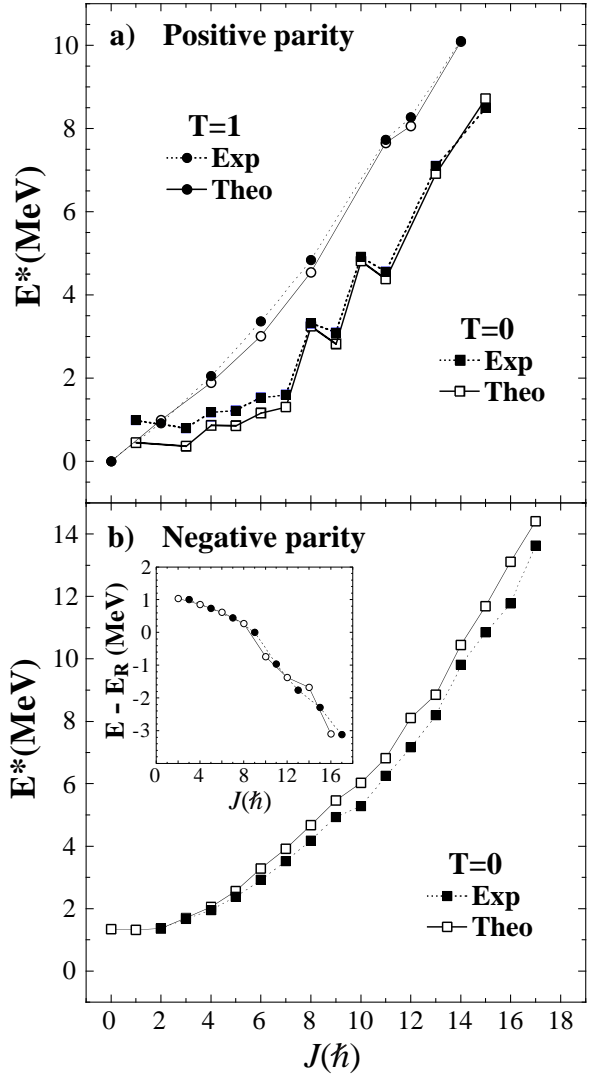}
\caption{Excitation energy versus angular momentum for states in $^{46}$V.
  The experimental data (filled symbols) are compared with shell model
  calculations. (a) The $T=0$ and $T=1$ positive parity bands. (b) The
  $T=0$ negative parity band; the inset shows the experimental
  excitation energy, minus an average rigid rotor contribution,
  plotted separately for each signature.}
\label{fig3}
\end{figure}

\begin{figure}
\epsfxsize=0.9\columnwidth
\epsffile{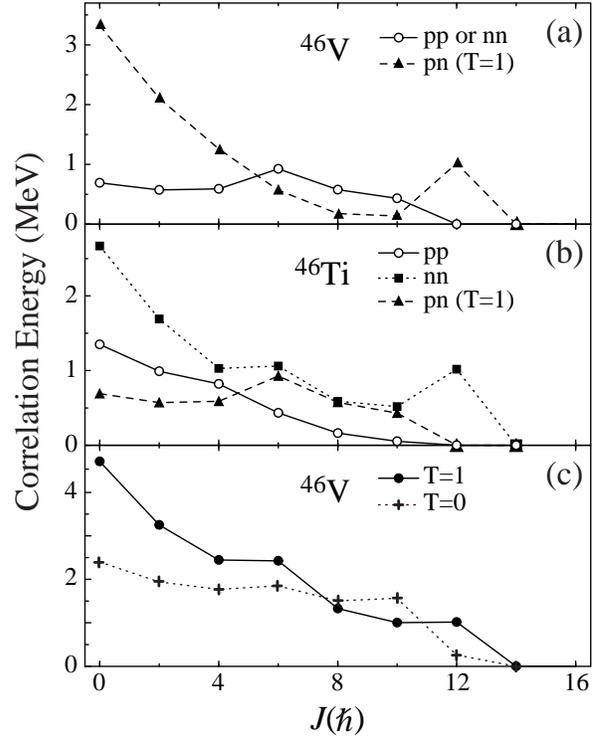}
\caption{Pairing correlation energy versus angular momentum for the $T=1$ 
  isobaric analogue states in $^{46}$V and $^{46}$Ti from SM
  calculations. (a) $T=1$ ($J=0$) pairing channels in $^{46}$V. (b)
  $T=1$ ($J=0$) pairing channels in $^{46}$Ti. (c) $T=1$ ($J=0$) and
  $T=0$ ($J=1$) pairing channels in $^{46}$V}.
\label{fig4}
\end{figure}

\begin{table}
\caption{Reduced matrix elements for transitions between 
  positive parity states in $^{46}$V from SM calculations.}
\label{table2}
\renewcommand{\arraystretch}{1.1}
\begin{tabular}{cccccc}
$I^{\pi}_i$ & $I^{\pi}_f$ & $B(M1)$  & $B(E2)$  & BR$_{theo}$  & BR$_{exp}$ \\  
~~~~~~~~ & ~~~~~~~~~~~ & $(\mu_N^2)$ & $(e^2fm^4)$ &  $(\%)$   & $(\%)$ \\
\hline
$1^{T=0}$ & $0^{T=1}$ &~~1.07  & ~~~~~~~  & ~~~~~~~~~~~ & ~~~~~~~~~\\
$2^{T=1}$ & $0^{T=1}$ &~~~~~~  & ~~~142  & ~~~~ &~~~~\\
$3^{T=0}_2$ & $2^{T=1}$ &~~0.15  & ~~~~~~~  & ~~~~~~~~~~~ & ~~~~~~~~~\\
$4^{T=1}$ & $2^{T=1}$ &~~~~~~  & ~~~187  & ~~~10 & ~~~~~ \\
~~~~~~~  & $3^{T=0}_2$ &~~0.63  & ~~~~~~~  & ~~~82 & ~~100 (20)\\
~~~~~~~  & $5^{T=0}_1$ &~~0.03  & ~~~~~~  & ~~~7  & ~~~~~ \\
$6^{T=1}$ & $5^{T=0}_2$ &~~0.77  & ~~~~~~~ &~~~42 & ~~~100 (40)  \\
~~~  & $5^{T=0}_1$ &~~~0.49  & ~~~~ & ~~~57 & ~~~~~~  \\
~~~  & $4^{T=1}$ &~~~~~~  & ~~~~175 & ~~~1 & ~~~~~~  \\
$8^{T=1}$ & $9^{T=0}_1$ &~~1.02  & ~~~~~~~ &~~~63 & ~~~~~  \\
~~~  & $7^{T=0}_2$ &~~~1.40  & ~~~~ & ~~28 & ~~~~~~ \\
~~~  & $7^{T=0}_1$ &~~~0.017  & ~~~~ & ~~~7 & ~~100 (45)  \\
~~~  & $6^{T=1}$ &~~~~~~  & ~~~~167 & ~~~2 & ~~~~~~  \\
$12^{T=1}$ & $13^{T=0}$ &~~2.56  & ~~~0.4 &~~21 & ~~~~  \\
~~~  & $11^{T=0}$ &~~~1.21  & ~~1.4 & ~~~78 & ~~100 (30)  \\
~~~  & $10^{T=1}$ &~~~~~~  & ~~~~54 & ~~~1 & ~~~~~~  \\
$14^{T=1}$ & $15^{T=0}$ &~~3.19  & ~~~~~~~ &~~~98 & ~~~100 (30)  \\
~~~  & $13^{T=0}$ &~~~0.048  & ~~~~ & ~~~1.5 & ~~~~~~  \\
~~~  & $12^{T=1}$ &~~~~~~  & ~~~~53 & ~~~0.5 & ~~~~~~  \\
\end{tabular}
\end{table}

\end{multicols}

\end{document}